\documentstyle[12pt]{article}

\newcommand{\gwig}{\mbox{\,\raisebox{.3ex}
{$>$}$\!\!\!\!\!$\raisebox{-.9ex}{$\sim$}}\,}

\font\tenrm=cmr10
\font\tenit=cmti10
\font\elevenbf=cmbx10 scaled\magstep 1
\font\elevenrm=cmr10 scaled\magstep 1
\font\elevenit=cmti10 scaled\magstep 1

\font\ninerm=cmr9

\font\sevenrm=cmr7
\font\fiverm=cmr5

\renewenvironment{thebibliography}[1]
{ \elevenrm
\begin{list}{\arabic{enumi}.}
{\usecounter{enumi} \setlength{\parsep}{0pt}
\setlength{\itemsep}{3pt} \settowidth{\labelwidth}{#1.}
\sloppy
}}{\end{list}}

\newcommand{\capfont}{\baselineskip=12pt \tenrm
\textfont0=\tenrm \scriptfont0=\sevenrm \scriptscriptfont0=\fiverm }

\begin{document}

\begin{titlepage}
\begin{flushright} CERN-TH.7005/93 \\ UCLA/93/TEP/32
\end{flushright}
\vspace{1ex}
\vspace{0.75cm}

\begin{center}
 {\elevenbf \vglue 10ptPHENOMENOLOGY OF MULTI-W PROCESSES\\
\vglue 3pt IN COSMIC RAYS}
\footnote{
\ninerm
\baselineskip=11pt
Invited talk presented by A. Ringwald at {\it
17th Johns Hopkins Workshop on Current Problems in Particle
Theory, Particles and the Universe}, July 30--August 1, 1993, Budapest,
Hungary}
\vspace{0.75cm}

{ {\bf D.A. Morris}\footnote{email: morris@uclahep.bitnet} \\
University of California, Los Angeles\\
405 Hilgard Ave., Los Angeles, CA,  U.S.A. \\
{\bf A. Ringwald}\footnote{email: ringwald@cernvm.cern.ch} \\
Theory Division, CERN\\
CH-1211 Geneva 23, Switzerland}
\date{}
\end{center}

\vspace{1cm}
\thispagestyle{empty}
\begin{abstract}\normalsize
\noindent
We report on a study of the potential of various cosmic ray
physics experiments to search for Standard Model processes
involving the nonperturba\-ti\-ve
production of $\gwig~{\cal O}(\alpha_W^{-1}) \simeq 30$
weak gauge bosons. Whereas present and near-future experiments are
insensitive to proton-induced processes,
neutrino-induced processes give rise to promising signatures and
rates in AMANDA, DUMAND, MACRO and NESTOR,
provided that
a cosmic neutrino flux exists at levels suggested
by recent models of active galactic nuclei.
The Fly's Eye currently
constrains the largest region of parameter space
characterizing multi-W phenomena.
\end{abstract}
\vspace{1.0cm}
\begin{flushleft} CERN-TH.7005/93 \\ UCLA/93/TEP/32 \\
September 1993
\end{flushleft}
\end{titlepage}

\newpage
\pagestyle{plain}
{\elevenbf\noindent 1. Introduction}
\vglue 0.4cm
\baselineskip=14pt
\elevenrm
A few years ago it was realized$^{1-4}$
that, even within the context of
a weakly coupled Standard Model,
lowest-order perturbative calculations of the
inelastic scattering of quarks and leptons
involving the production
of $\gwig~{\cal O}(\alpha_W^{-1}) \simeq 30$
weak gauge bosons result in an explosive (and unitarity violating)
growth of the associated parton-parton cross section above
center-of-mass energies
$ \gwig~ {\cal O}(\alpha_W^{-1} M_W ) \simeq 2.4$~TeV.
Thus, perturbation theory, when
applied to large order processes ({\it e.g.}, like the production of
${\cal O}(\alpha_W^{-1})$ weak bosons) breaks down somewhere
in the multi-TeV range.
It is presently an open theoretical
question whether large-order weak interactions
become strong at this energy scale
(in the sense of having observable cross sections)
or whether they remain unobservably small at
all energies. The answer almost certainly lies beyond the realms
of conventional perturbative techniques
(see Ref.~5 for an overview).
Given the stakes involved, a quantitative
consideration of experimental constraints on
multi-W production is clearly desirable.
In this talk we
report on our recent investigations of this question$^{6,7}.$

Until the commissioning
of the proposed Large Hadron Collider (LHC) and the
Superconducting Super Collider (SSC), which
would be ideal for
observing or constraining multi-W phenomena over a wide range of
energies and cross sections$^{8,9}$,
cosmic rays provide the only access to the required energy
scales. In order to investigate multi-W processes in a quantitative
manner we take a phenomenological approach.
We first parametrize  multi-W processes in terms of parton-level
thresholds and cross sections and then explore the sensitivity of
current and near-future cosmic ray experiments to various signatures
of multi-W phenomena. We consider both
atmospheric and underground phenomena
induced by cosmic ray protons
and neutrinos.

We demonstrate that even if
proton-induced atmospheric multi-W phenomena occur in Nature,
the features of the resulting air showers are unlikely to allow
one to distinguish them from fluctuations
in a much larger background of generic showers.
On the other hand, ultrahigh energy
neutrinos, for which a sizeable flux has recently
been conjectured from sources such
as active galactic nuclei$^{10-13}$
offer exciting
possibilities for observing or constraining multi-W phenomena$^{14,15}.$
Subsurface detectors such
as AMANDA$^{16}$,
DUMAND$^{17}$,
MACRO$^{18}$,
and
NESTOR$^{19}$
can be sensitive to neutrino-induced multi-W phenomena
in certain regions of multi-W parameter space.

The structure of this talk is as follows. In sect.~2
we characterize multi-W phenomena by a two-parameter
working hypothesis which frees us from specifying an
underlying (most likely nonperturbative) mechanism for
multi-W production. We also describe the
gross features of multi-W phenomena and present
discovery limits for the LHC and SSC.
We discuss proton-induced and neutrino-induced multi-W phenomena
in sects.~3 and 4, respectively.
We consider a variety of detection techniques and present
discovery limits which may be contrasted with the
superior sensitivity of future hadron colliders. In sect.~5 we
conclude.
\newpage
\elevenbf\noindent 2. Multi-W Parameterization
\vglue 0.4cm
\baselineskip=14pt
\elevenrm
In the absence of a reliable first-principles calculation
of multi-W production, we
parametrize
the quark-quark or neutrino-quark
cross section for multi-W production by
\begin{equation}
\label{eq:workhyp}
\hat \sigma_{\rm multi-W} = \hat \sigma_0\
\Theta
\bigl( \sqrt{\hat s} -\sqrt{\hat s_0 } \bigr) .
\end{equation}
By convoluting the subprocess cross section of Eq.~(\ref{eq:workhyp})
with the appropriate quark distribution functions, one obtains the
multi-W production cross sections of Fig.~1
for protons and neutrinos
striking stationary nucleons and electrons. Motivated by suggestions that
multi-W production may be an intrinsically nonperturba\-tive
phenomenon, the
quark distribution functions are evaluated at a momentum transfer scale
of $Q^2 = M_W^2.$
The curves in Fig. 1 are universal in the sense that they
have been scaled by $\hat \sigma_0$ and $\sqrt{ \hat s_0 }$.

\begin{center}
\hspace*{0in}

\parbox{12cm}{ \capfont

Fig. 1. Universal curves parametrizing multi-W production cross sections
in proton-nucleon ($\capfont p N$), proton-electron ($p e^-$),
neutrino-nucleon ($\capfont \nu N$) \linebreak and
neutrino-electron ($\capfont \nu e^-$)
collisions. Curves are for protons and neutrinos with
laboratory energy $\capfont E$ colliding with
nucleons and electrons at rest. $\capfont E_p^{(pN~{\rm thresh})}
= \hat{s}_0 / ( 2 m_p)$ is the proton threshold energy
for $\capfont pN$ multi-W processes. }
\end{center}

For definiteness, we will assume
throughout this talk that $\hat\sigma_0$ refers to the production
of exactly 30 W bosons; allowing for the production of variable numbers
of W's (and Z's and possibly prompt photons)
is straightforward but is an unnecessary complication
at the level of our investigation.
An optimistic range of parameters to consider might encompass
\begin{equation}
\begin{array}{lcccl}
\label{eq:threshold}
\displaystyle{m_W \over \alpha_W } \simeq 2.4~\mbox{TeV} &
\leq &
\sqrt{\hat s_0} &
\leq &
40~\mbox{TeV}, \\
 & & & & \\
\label{eq:partoncro}
\displaystyle{\alpha_W^2 \over m_W^2 } \simeq 100 ~\mbox{pb} &
\leq &
\hat \sigma_0  &
\leq &
\sigma^{pp}_{\rm inel} \times
\left( \displaystyle{1~\mbox{GeV}\over m_W}\right)^2 \simeq
10~\mu\mbox{b}.
\end{array}
\end{equation}
The lower limit of $\sqrt{\hat s_0}$ is suggested by the
energy scale at which perturbation theory becomes unreliable$^{3,4}$
whereas the upper range is of the order of
the sphaleron mass$^{20}.$
The lower limit of
$\hat \sigma_0 $ is characteristic of
a geometrical ``weak'' cross section and the upper limit
range is a geometrical ``strong''
cross section suggested by analogies between the weak
SU(2) gauge sector and the color SU(3) gauge sector$^{21}.$

The simultaneous production of ${\cal O}(30)$ W~bosons
at future hadron colliders like the LHC or SSC would lead to
spectacular signatures$^{8,9}.$
Since  approximately 20  charged hadrons (mainly $\pi^\pm$'s) arise
from hadronic W decays one
could typically expect
400 $\pi^\pm$'s
in one multi-W event accompanied by $\simeq 400$ photons from the
decay of $\simeq 200$ $\pi^0$'s. The charged hadrons would
have a minimum average transverse momentum of order
$p_T^\pi \ge {\cal O}
(m_W/30 )\simeq (2-3)$ GeV
if the W bosons are produced without transverse momentum.
Similarly, one could expect $\simeq 5$
prompt muons ($\simeq 3$ from W decays
and $\simeq 2$ from $c$, $b$, or $\tau$ decay)
carrying a minimum average transverse momentum of
$p_T^\mu \ge {\cal O}( m_W/2 )\simeq 40$ GeV.
Analagous situations hold for other prompt leptons such as $e^\pm,
\nu$ etc.
No conventional reactions in the Standard Model are backgrounds to
multi-W processes$^{9}.$

Figure 2
shows the regions in $\sqrt{\hat{s}_0} - \hat \sigma _0$
space accessible to the LHC
and the SSC.
The contours
correspond to 1 and 10 events (assuming 100\% detection efficiency)
for $10^7$ s of operation. These contours may be used
as a benchmark to evaluate the effectiveness of various cosmic
ray physics experiments for constraining multi-W phenomena.
\vglue 0.6cm
\elevenbf\noindent 3. Proton-Induced Multi-W Processes
\vglue 0.4cm
\baselineskip=14pt
\elevenrm
Cosmic ray protons and heavy nuclei constitute a guaranteed
flux of high-energy primaries potentially capable of initiating
multi-W phenomena. In this section we briefly review$^7$
the possibility
of exploiting this cosmic flux and isolating multi-W phenomena from
generic hadronic reactions at conventional surface arrays
which reconstruct the features of an extensive
air shower by interpolating or extrapolating measurements
of a shower's particle content.
We restrict our attention to
cosmic ray protons since they provide the dominant flux in terms
of energy per nucleon.

Multi-W phenomena initiated by cosmic ray {\it protons} are plagued
by {\it small rates}
and {\it poor signatures} due to competing purely hadronic
processes with ${\cal O}$(100~mb) cross sections.
The small rates are demonstrated in
Figure 3 which
shows contours
\begin{center}
\hspace*{0in}

\parbox{12cm}{ \capfont
Fig. 2. Contours corresponding to 1 and 10 multi-W events
in one year (10$^7$ s) of
operation for the LHC
($\sqrt{s}=14.6~\mbox{TeV};
{\cal L}=10^{34}~\mbox{cm}^{-2}\mbox{s}^{-1}$)
and the SSC
($\sqrt{s}=40~\mbox{TeV};
{\cal L}=10^{33}~\mbox{cm}^{-2}~\mbox{s}^{-1})$.
\label{fi:lhcssc}
  }
\hspace*{0in}

\parbox{12cm}{ \capfont Fig. 3.
Event number contours in $10^7$~s
for proton-induced multi-W air showers
assuming the Constant Mass Composition model for proton flux.
Solid: 100 km$^2$ conventional surface array sensitive to
$E_{\rm shower} \ge 1$ PeV at zenith angles
$\theta \le 60^{\rm o}$. Dashed: Fly's Eye array sensitive to
$E_{\rm shower} \ge 100$~PeV using aperture of Ref.~30.
  }
\end{center}
for the number of
proton-induced multi-W air showers at zenith angles
$\theta \le 60^{\rm o}$ striking a 100~km$^2$ conventional
surface array in $10^7~\mbox{s}.$ For our calculations we
use the cosmic proton flux of the Constant Mass Composition (CMC)
model$^{22}$
(see Fig. 4).
For purposes of illustration we
optimistically assume an array threshold energy of
$E_{\rm thresh} = 1$~PeV
which accommodates
all multi-W thresholds above $\sqrt{\hat{s}_0} \ge 2.4$ TeV.
In 100~km$^2$ arrays like AGASA$^{23}$
and EAS-100$^{24}$
inter-detector spacing on the order of .5--1~km makes
$E_{\rm thresh}=100-1000$~PeV more realistic but
does not change our conclusions.

\begin{center}
\hspace*{0in}

\parbox{12cm}{ \capfont Fig. 4.
Differential flux of protons and neutrinos
used in text. The Constant Mass Composition proton flux
is from Ref.~22. The diffuse
neutrino flux from active galactic nuclei (AGN) and
the 2.7 K photoproduced neutrino flux are taken from
Ref.~10.
Neutrino fluxes shown are
summed over species in the proportion $\nu_\mu:
\bar\nu_\mu:\nu_e:\bar\nu_e = 2:2:1:1$.
\label{fi:pflux}
}
\end{center}

Though the region of the ($\sqrt{\hat{s}_0},\hat{\sigma}_0$) plane
accessible to proton-induced multi-W phenomena is small, let
us now investigate whether
it is possible to convincingly distinguish
an air shower of multi-W origin from
a generic hadronic air shower.
For the remainder of this section, we restrict our
attention to the optimistic scenario of
a parton-parton multi-W threshold of $\sqrt{\hat{s}_0} = 2.4$~TeV
with $\hat{\sigma}_0 = 10~\mu\mbox{b}$. For this choice of parameters
the most probable shower energy is $\simeq 30$ PeV whereas the
average shower energy is $\simeq 250~\mbox{PeV}$ due to a long tail
on the corresponding distribution.

Consider the characteristics of the most probable ( $\simeq 30$~PeV)
multi-W air showers.
To phrase our results in experimentally relevant terms
we use the computer program {\sc showersim}$^{25}$
to simulate multi-W air showers and generate samples
of generic proton-induced and iron-induced showers.
Figure 5
compares 30~PeV multi-W, proton and iron showers
in terms of radial particle densities
(with respect to a vertical shower axis)
of electrons ($E_e \ge 1~\mbox{MeV}$),
muons $(E_\mu \ge 1~\mbox{GeV}$)
and hadrons ($E_{\rm had} \ge 1~\mbox{GeV}$).
Each curve is averaged over 25--100 showers taking into
account the distribution of the depth of first interaction
in the upper atmosphere. The densities
in Fig.~5 correspond to an
an observation depth of $800~\mbox{g/cm}^2$
(roughly the CYGNUS array depth$^{26}$
).

\begin{center}
\hspace*{0in}

\parbox{12cm}{ \capfont Fig. 5.
Lateral distributions of electrons,
muons and hadrons in 30~PeV vertical air showers
at atmospheric depth of 800 g/cm$^2.$ Solid curves correspond to
proton-induced multi-W showers assuming $\sqrt{\hat s_0}= 2.4$ TeV
and any value of $\hat\sigma_0.$
Dashed (dot-dashed) curves correspond to generic showers initiated by
proton (iron) primaries. Each curve is an average over 25--100 showers
including variations in the depth of first interaction.
 }
\end{center}

The differences between the particle density profiles of 30~PeV showers
in Fig.~5 are hardly dramatic.
While there are identifiable systematic differences
between average showers of different origin, the differences
do not appear to be sufficient to discriminate between multi-W
showers and fluctuations in generic proton or iron showers.
We emphasize this point by noting that in the CMC flux model,
the differential fluxes of 30~PeV generic proton-induced,
iron-induced and multi-W showers (with
$\sqrt{\hat{s}_0} = 2.4$~TeV, $\hat{\sigma}_0 = 10~\mu\mbox{b}$)
stand in the proportion p : Fe : multi-W
$ \simeq 1.2 \times 10^5 : 1.1 \times 10^5 : 1.$
Therefore, the prospects for
detecting proton-induced multi-W phenomena using conventional
surface arrays are poor (see Ref. 7
for further details).

\vglue 0.5cm
\elevenbf\noindent 4. Neutrino-Induced Multi-W Processes
\vglue 0.3cm
\baselineskip=14pt
\elevenrm
Ultrahigh energy neutrinos could provide particularly striking
signatures for multi-W processes. In contrast to proton-induced
multi-W processes which must compete with ${\cal O}(100$~mb)
generic hadronic processes, neutrino-induced phenomena
must contend only with ${\cal O}$(nb) weak interaction processes.
Even if the total neutrino-nucleon multi-W cross section
is as large as ${\cal O}(\mu$b), the majority of all interactions
occur deep in the atmosphere or inside the Earth.

A precondition for observing or constraining
neutrino-induced multi-W phenomena is, of course,
the existence of a flux of ultrahigh energy neutrinos.
Though atmospheric neutrinos, {\it i.e.,} neutrinos produced in hadronic
showers in the atmosphere, provide a guaranteed
diffuse flux of ultrahigh energy neutrinos, their overall
rate in the PeV region is anticipated to be negligible
(see, {\it e.g.}, Ref.~13
).
Much more promising are the recent predictions
of a sizeable flux of PeV neutrinos from active galactic
nuclei (AGN)$^{10-13}.$
In this section we discuss the implications
of a large AGN neutrino flux for multi-W processes. For definiteness,
we use the (revised) Stecker {\it et al.}$^{10}$ AGN neutrino flux.
AGN neutrino fluxes calculated under different assumptions
in Refs.~11,12
generally agree with Ref.~10
above .1~PeV, which is the energy range we are interested in.
In this sense our use of the Stecker {\it et al.} flux
is intended to be representative of a large class
of AGN flux models. In Ref. 7
we have checked that,
within the parameter ranges of Eq.~(2), large neutrino
cross sections for multi-W production are consistent with
proposed AGN flux models.

In addition to proposed AGN neutrino fluxes, we also consider
the possibility of detecting neutrinos which are photoproduced by
protons scattering inelastically off the 2.7~K cosmic background
radiation (CBR)$^{27}$, producing charged pions which subsequently
decay to neutrinos$^{28,29}.$ As shown in Fig.~4,
such processes may provide the
dominant component of the neutrino flux at
energies beyond $\simeq 1$ EeV.
The photoproduced neutrino flux,
$j_\nu^{\rm 2.7~K},$ shown in Fig.~4
is taken from Ref. 10.

We divide our discussion of neutrino-induced multi-W phenomena
into two sections.
In sect.~4.1 we discuss constraints on neutrino-induced
multi-W production from the Fly's Eye.
In sect.~4.2 we evaluate the potential of subsurface detectors
to observe the showers associated with multi-W phenomena
and discuss the detection of
distant multi-W phenomena through
searches for energetic muon bundles.

{\elevenit\noindent 4.1 The Fly's Eye Limits}
\vglue 0.4cm
\baselineskip=14pt
\elevenrm
The Fly's Eye$^{30}$
is  an optical array sensitive to
nitrogen fluorescence light from air showers whose
trajectories do not necessarily intersect the array$^{31}.$
By detecting fluorescence light emitted as air showers
streak across the sky, the
Fly's Eye is capable of reconstructing the longitudinal
development of air showers with energy greater than
$E_{\rm thresh}= 100$~PeV.

Independent of any neutrino flux model,
the Fly's Eye array
puts upper limits$^{32}$ on
the product of the flux times total cross section
for weakly interacting particles
in the range $10^8~\mbox{GeV} \leq E_\nu \leq 10^{11}~\mbox{GeV}$
assuming that such particles initiate
extensive air showers deep in the atmosphere.
The limits are deduced from the non-observation of downward-moving
air showers within the Fly's Eye fiducial volume such that the shower
axis is inclined
80$^{\rm o}$ to 90$^{\rm o}$ from the zenith at the point of
impact
on the Earth. Showers meeting these criteria could only have been
initiated by particles typically penetrating more than
3000 g/cm$^2$ of atmosphere before interacting, which excludes
showers initiated by ultrahigh energy photons and hadrons.

Assuming that the weakly interacting particles referred to by the
Fly's Eye are neutrinos, we denote the relevant cross section
by $\sigma_{\rm tot}^{\nu N}(E_\nu)$ which receives contributions from
both multi-W and familiar charged current weak interactions.
The Fly's Eye limits may be summarized$^{14,33}$ by
$(j_\nu \sigma_{\rm tot}^{\nu N})_{\rm Fly's~Eye}$
$\leq 3.74 \times 10^{-42}
\, \times \,
( E_\nu / 1$ GeV)$^{-1.48}$ s$^{-1}$ sr$^{-1}$ GeV$^{-1}.$
Since these limits neglect the possibility of flux attenuation
in the upper atmosphere due to large inelastic cross sections,
they nominally apply only if
$\sigma_{\rm tot}^{\nu N}(E_\nu) \le 10~\mu$b.
If one considers a particular neutrino flux model
$j_\nu^{\rm model}(E_\nu)$
the Fly's Eye limit excludes regions
in the $(E_\nu,\sigma_{\rm tot}^{\nu N})$ plane bounded by
\begin{equation}
\label{eq:fe}
\begin{array}{lcccl}
\displaystyle{
( j_\nu \sigma_{\rm tot}^{\nu N})_{\rm Fly's~Eye}
               \over
j_\nu^{\rm model}}
 &  < & \sigma_{\rm tot}^{\nu N}(E_\nu) & < & 10~\mu{\mbox{b}}\, , \\
  & & & & \\
10^8~{\mbox{GeV}} &  < &  E_\nu & < &   10^{11}~{\mbox{GeV}}  .
\end{array}
\end{equation}

These inequalities exclude  a corresponding
region in $(\sqrt{{\hat s}_0},{\hat \sigma}_0)$ space which
parametrizes multi-W phenomena.
Figure 6 shows the excluded (hatched) region
of multi-W parameter space for the AGN neutrino flux of
Stecker {\it et al.} ({\it i.e.,}
$j_\nu^{\rm model} = j_\nu^{{\rm Stecker}~et~al.} $ from Fig.~4).
If one takes $j_\nu^{\rm model} = j_\nu^{{\rm Stecker}~et~al.} +
j_\nu^{\rm 2.7~K}$ in Eq.~(\ref{eq:fe}), the Fly's Eye limit
 enlarges the
excluded region in $(\sqrt{\hat s_0},\hat\sigma_0)$ space
by the area labelled ``2.7~K Photoproduced $\nu$'' in
Fig.~6 (double hatched).

Though the appearance of an enlarged excluded region
is welcome, it is sensitive to details of the assumed CBR flux.
Had we assumed a CBR flux component $j_\nu^{\rm 2.7~K}$
which was a factor of ten smaller than that shown in Fig.~4.
(corresponding to a lower redshift),
the Fly's Eye limit would not have
introduced an additional constraint$^7$. We should keep such
uncertainties in mind to avoid attaching undue
significance to the excluded regions in
Fig.~6.
Nevertheless it is intriguing to speculate about
detecting CBR neutrinos via multi-W processes since
the prospects
\begin{center}
\hspace*{0in}

\parbox{12cm}{ \capfont Fig. 6.
Regions of multi-W parameter space excluded by the Fly's Eye.
The region labelled ``AGN $\nu$'' is excluded if one assumes
only the AGN neutrino flux of Stecker \mbox{\tenit et al.}$^{10}.$
The region labelled ``2.7 Photoproduced $\nu$'' is
excluded in addition if one includes the neutrino flux
contributions due to the inelastic scattering of protons off the
cosmic background radiation
shown in Fig.~4.
\label{fi:neutrinoinitiatedflyseyeexcluded}
}
\end{center}
for detecting such neutrinos through generic
weak interactions is poor unless the CBR neutrino
flux is associated with a very large redshift.

\vglue 0.6cm
{\elevenit\noindent 4.2 Subsurface experiments}
\vglue 0.4cm
\baselineskip=14pt
\elevenrm
Detectors deep below the surface of the Earth, be they
shielded by rock ({\it e.g.} MACRO$^{18}),$
water ({\it e.g.} DUMAND$^{17}$,
NES\-TOR$^{19},$
Baikal NT-200$^{34}$ )
or ice (AMANDA$^{16}$)
offer a unique perspective on neutrino-induced phenomena, as has been
nicely reviewed by Francis Halzen
and Leo Resvanis during this workshop. In this
section we investigate two possible modes for detecting
neutrino-induced multi-W phenomena using subsurface experiments.
We first consider the prospects
for
observing contained neutrino-induced multi-W phenomena
and later turn to the
detection of muon bundles arising from neutrino interactions
in the surrounding medium.

     Aside from the energy involved, contained
neutrino-induced multi-W production would reveal its
origins by its enormous multiplicity
(${\cal O}$(400) charged hadrons, ${\cal O}(400)$ photons,
and a few prompt muons and electrons).
Generic deep inelastic $\nu N$ scattering and the
resonant process $\bar\nu_e + e^- \rightarrow W^-
\rightarrow {\rm hadrons}$ can also give contained hadron
production, but only with significantly lower multiplicity.
Figure 7 shows contours for contained multi-W events in $10^7$~s in
a .2~km$^3$  volume of water at an ocean depth of 4.5 km (approximately
the characteristics of DUMAND for PeV contained showers$^{35}$).

\begin{center}
\hspace*{0in}

\parbox{12cm}{\capfont Fig. 7.
Contours for neutrino-induced contained multi-W events in .2~km$^3$
volume of water at an ocean depth of 4.5 km in 10$^7$ s (appropriate
for contained PeV showers
in DUMAND). The neutrino flux of Stecker \mbox{\tenit et al.}$^{10}$
is assumed (see Fig. 4).}
\end{center}

Of particular interest is the ability for
subsurface detectors to detect muons which arise from
energetic neutrino interactions up to a few kilometers
away. For distant multi-W production the effects of producing
hundreds of hadrons and photons will have died off well before
reaching the detector but the anticipated 2--3 muons from prompt
W decays produced with $E_{\mu} \simeq {\cal O}(100$ TeV)
and $p_T^\mu \simeq {\cal O}(40$ GeV) propagate great distances.
The signature of multi-W production in this case would be
energetic muon bundles.

The ability to detect muons from distant neutrino reactions
increases a subsurface detector's effective neutrino target
volume dramatically and is the premise upon which such
detectors can act as neutrino telescopes.
Considerable effort has recently been directed towards
the prospects of detecting ultrahigh energy neutrinos (most
likely from AGN) using subsurface detectors$^{13}.$
Despite their limited sensitivity to such phenomena,
Fr\'ejus$^{36}$
and Soudan--2$^{37}$
have already placed useful observational constraints
on AGN flux models.

As discussed in Ref.~14,
near-horizontal muon bundles in DUMAND and MACRO would
be characteristic of neutrino-induced multi-W phenomena. By concentrating
on large zenith angles, one can avoid the complications from
a large background of muon bundles from generic hadronic
interactions in the atmosphere.
We present in Fig.~8 contours
for muon bundles  beyond zenith angles of $80^{\rm o}$ for
MACRO and DUMAND for the AGN neutrino flux of Stecker
{\it et al.}$^{10}$;
the contours expected for AMANDA and NESTOR (stage 1) are similar
to those of the DUMAND contours.
Due to our assumed production of 30 W bosons, each muon bundle
consists of approximately 3 muons. The average muon energy
$\langle E_\mu \rangle$
entering the detector and
the average inter-muon separation $\langle r_\mu\rangle$
depend on $\sqrt{\hat s_0}$ and $\hat\sigma_0.$
For
example, for $(\sqrt{\hat s_0} = 4$ TeV,
$\hat\sigma_0=10$ nb)
one expects $\simeq 1.5$ bundles per $10^7$~s in DUMAND
with $\langle E_\mu \rangle \simeq {\cal O}(180$~TeV)
and
$\langle r_\mu \rangle = {\cal O}(2.5$ m);
for $(\sqrt{\hat s_0} = 4$ TeV, $\hat\sigma_0=1~\mu$b) one
expects $\simeq 30$ bundles per $10^7$~s in DUMAND
with $\langle E_\mu \rangle \simeq {\cal O}(70$ TeV)
and $\langle r_\mu \rangle = {\cal O}(3.6$ m).
Assuming an additional 2.7~K photoproduced neutrino flux
component at the level shown in Fig.~4
changes the contours of Fig.~8
by a negligible amount.
\begin{center}
\hspace*{0in}

\parbox{12cm}{ \capfont Fig. 8.
Contours for neutrino-induced multi-W muon
bundles at zenith angles $\theta > 80^{\rm o}$ in 10$^7$ s
at MACRO and DUMAND assuming the AGN neutrino flux of
Stecker \mbox{\tenit et al.}$^{10}$. }
\end{center}

It may also be possible to constrain multi-W phenomena
by searching for non-horizontal muon bundles and thereby enlarge
the accessible region in $\sqrt{\hat s_0}$--$\hat \sigma_0$ space.
Fig. 9 shows contours for
muon bundles for zenith angles between 0$^{\rm o}$ and 180$^{\rm o}$ for
MACRO for the Stecker {\it et al.} AGN neutrino flux.
An additional 2.7~K photoproduced neutrino flux component
at the level of fig.~4
has a negligible effect on the
MACRO contours.

\begin{center}
\hspace*{0in}

\parbox{12cm}{ \capfont Fig. 9.
Contours for neutrino-induced multi-W muon
bundles for all zenith angles in 10$^7$ s
at MACRO assuming the AGN neutrino flux of
Stecker {\it et al.}$^{10}.$
Preliminary rates for muon bundles in MACRO
presented in Ref.~6
included only AGN $\nu_\mu-$induced multi-W
processes and hence are smaller than those shown above
by a factor of 3. }
\end{center}

Whereas the inter-muon separation
expected from generic hadronic interactions high in the atmosphere
is typically of ${\cal O}(5$--10 m), neutrino-induced multi-W phenomena
occur primarily inside the Earth and  result in much more spatially
compact muon bundles.
Figure 10 compares MACRO data for
pair-wise muon separation to the contribution expected from
neutrino-induced multi-W phenomena for $(\sqrt{\hat s_0}=2.4$ TeV,
$\hat\sigma_0 = 10~\mu$b). The MACRO
data is taken from Fig.~4 of Ref.~38
and corresponds to
muon bundles at zenith angles $\theta < 60^{\rm o}$
detected by two supermodules
operating for 2334.3 hours.
The MACRO data contains contributions from muon bundles of
all multiplicities; approximately half of the reconstructed
pairs come from $n_\mu =2$ muon bundles. We suggest that by
separately examining the pair-wise muon separation in
bundles with fixed numbers of muons ({\it e.g.,} $n_\mu$=3)
\begin{center}
\hspace*{0in}

\parbox{12cm}{ \capfont Fig. 10.
MACRO pairwise muon separation data$^{38}$
compared with
expectations for neutrino-induced multi-W phenomena
for $(\sqrt{\hat s_0} = 2.4$ TeV , $\hat\sigma_0 = 10~\mu$b)
assuming the AGN neutrino flux of Stecker \mbox{\tenit et al.}$^{10}.$
MACRO data corresponds to two supermodules operating for 2334.3 hours
sensitive to bundles with zenith angle $\theta < 60^{\rm o}.$
 }
\end{center}
as has been done by the Fr\'ejus collaboration$^{39},$
MACRO may be able to put constraints on the existence of
multi-W phenomena. A particularly useful signature of multi-W processes
in this respect is the energy carried by each muon. Muons
arising from multi-W processes in Fig.~10 would
have energies of approximately 80~{\rm TeV} as they enter the detector
and may be distinguished by mechanisms such as catastrophic energy
loss$^{36,37}.$
Though some of the region
in $(\sqrt{\hat{s}_0},\hat\sigma_0)$ space to which MACRO is sensitive
is already excluded by the Fly's Eye (assuming the same AGN neutrino
flux), valuable independent constraints may already be possible from
existing MACRO data.
\vglue 0.6cm
\elevenbf\noindent  5. Conclusions
\vglue 0.4cm
\baselineskip=14pt
\elevenrm
\hspace*{\parindent}
The short term outlook for constraining or detecting
multi-W phenomena in cosmic
ray  physics  is  mixed.
Without  making additional  assumptions
(such as assuming the existence of a large cosmic neutrino flux)
one must focus on
proton-induced processes and
conclude that current and future experiments are
effectively insensitive to multi-W phenomena
over the entire range of parameter space where they
might plausibly exist. From this viewpoint one must
wait for terrestrial supercolliders before conclusive
constraints on multi-W processes are established.

An intermediate
scenario is one in which a flux of ultrahigh energy neutrinos
is detected in the future but is found to have interactions
consistent with generic charged current processes. This may
may place valuable constraints on the existence of multi-W phenomena,
notably from the Fly's Eye limits.

The most optimistic scenario requires a sizeable diffuse flux of
ultrahigh energy neutrinos. In this case
dedicated subsurface detectors such as
AMANDA, DUMAND, MACRO and NESTOR may even indicate
whether multi-W processes are real or an artifact of our
imperfect understanding of multi-TeV weak interactions.

\vglue 0.6cm
\elevenbf\noindent  6. Acknowledgements
\vglue 0.4cm
\baselineskip=14pt
\elevenrm
A.R. thanks the Organizing Committee for hospitality at
this very stimulating workshop. D.A.M. is supported by the Eloisatron project.
\vglue 0.6cm
\elevenbf\noindent  7. References
\vglue 0.4cm
\baselineskip=14pt
\elevenrm

\end{document}